\documentclass[aps,prb,twocolumn,showpacs]{revtex4}

\begin{document}
\title{Magnetic field dependence of pairing interaction  in  ferromagnetic superconductors with triplet pairing}

\author{V. P. Mineev}

\affiliation{Commissariat \`a l'Energie Atomique,
INAC/SPSMS, 38054 Grenoble, France}
\date{\today}

\begin{abstract}
It is developed a microscopic description of superconductivity in ferromagnetic materials with triplet pairing triggered by the exchange of magnetic fluctuations. Instead widely used    paramagnon model 
we work with phenomenological spectrum of fluctuations in the orthorhombic ferromagnet with strong magnetic anisotropy.  Depending of the  field orientation parallel or perpendicular to the direction of spontaneous magnetization the effective  amplitude of pairing interaction proves to be decreasing or increasing function of magnetic field that allows to explain the drastic difference in magnitudes of upper critical field in these directions.

\end{abstract}


\maketitle

\section{Introduction}

Superconductivity in ferromagnetic state found in  uranium compounds URhGe and UCoGe
\cite{Aoki01,Huy07} exhibits peculiar superconducting properties 
 such as extremely high upper critical field 
 \cite{Hardy05,Huy08,Slooten,Aoki09}  and the reentrant superconductivity in large external magnetic fields\cite{Levy05,Levy07}. These  ferromagnets have orthorhombic structure with magnetic moment oriented along $\hat c$ axis. 
  The observed
  upper critical field in the directions perpendicular to the magnetization
  proves  much higher than the paramagnetic limiting field. This unexpected for the superconductors with triplet pairing behavior \cite{Book,Choi} is explained by the band splitting caused by exchange interaction \cite{Min10}.
On the other hand, the paramagnetic limitation in a superconductor with equal spin pairing for the field parallel to the spin quantization direction  is  absent \cite{Book}. However, experimentally,
the upper critical field in the direction of spontaneous magnetization approximately coincides with the paramagnetic limiting field in these materials. The question, why the upper critical field parallel to $\hat c$ axis is so much smaller than the $H_{c2}$ in the perpendicular directions, remains open.

Another peculiar feature observed in UCoGe \cite{Huy08,Slooten,Aoki09} for the field directed perpendicular to the spontaneous magnetization is a pronounced upward curvature. This fact cannot be explained by the field dependent increase of the effective mass probably responsible for the S-shape upper critical field along b axis behavior   in the field region from 5 to 15 Tesla. Indeed, the upward curvature of  $H_{c2}(T)$ along a and b directions is observed already at much smaller fields  where effective mass falls down with field increase\cite{Aoki09} stimulating the opposite that is downward curvature tendency in the $H_{c2}(T)$ behavior.
The possible explanation for anomalous upper critical field curvature pointed out in the paper \cite{Huy08} is attributed to the crossover between two phases in a two-band ferromagnetic superconductors \cite{Cham,Min04}. 

Here we discuss the circle of problems related to the upper critical field behavior specific for superconducting ferromagnets URhGe and UCoGe. Our treatment is based on semi-microscopic description of superconductivity in ferromagnetic materials with the Cooper pairing caused by the magnetic fluctuations. Instead of wide spread paramagnon or Fermi liquid approach to the 
nonphonon mechanism of superconductivity (see for instance \cite{Fujimoto} and references therein) we shall use the phenomenological description of magnetic fluctuations in the orthorhombic ferromagnet with strong spin-orbital coupling. Certainly the complete theory of triplet superconductivity in ferromagnets has to take into account the frequency dependence of the pairing interaction. The invalidity of the Migdal theorem adds, however, the supplementary troubles to the Eliashberg type theory of superconductivity in heavy fermionic materials. Leaving this problem for the future investigations \cite{Tada}  we shall see to what kind of qualitative conclusions one can  come in frame of semi-microscopic weak coupling theory. 

To investigate the effect of the pairing interaction field dependence we  drop out all the orbital effects.  In other words we will solve the problem of the critical temperature field dependence
as if the  magnetic field acts only on the electron spins. This simplification allows to demonstrate the pure effect of suppression of superconducting state for the field parallel to magnetization and the opposite effect of stimulation of superconductivity for the field directed in perpendicular to magnetization directions. 

It should be noted that the proposed mechanism of pairing interaction field dependence is not an alternative to the discussed in literature mechanism related to the effective mass field dependence \cite{Aoki09,Miyake08,Miyake09,Aoki11}. Both of them make a contribution to the interplay of magnetism and superconductivity in ferromagnetic compounds. The unified treatment taking into account both mechanisms  in frame of some general approach is the  problem for the future.

The paper is organized as follows.   The critical temperature field dependence for  field orientations parallel and perpendicular to the direction of spontaneous magnetization are derived in two separate sections of the paper following each other. Both of them are based on the pairing interaction expressed through the components of susceptibility tensor derived in the Appendix.

\section{Pairing Hamiltonian} 

Following the paper \cite{Sam} we introduce the  interaction between electrons due to the spin fluctuations as
\begin{equation}
H_{int}=\frac{1}{2}\sum_{{\bf k}{\bf k}'{\bf q}}D_{ij}({\bf k},{\bf k}')\sigma^i_{\alpha\delta}\sigma^j_{\beta\gamma}a^\dagger_{{\bf k}\alpha} a^\dagger_{-{\bf k}\beta} a_{-{\bf k}'\gamma}a_{{\bf k}'\delta}.
\end{equation}
The matrix $D_{ij}$ is taken as the part of static spin susceptibility being
an odd function of both of its arguments 
$$
D_{ij}(-{\bf k},{\bf k}')=D_{ij}({\bf k},-{\bf k}')=-D_{ij}({\bf k},{\bf k}').
$$
It determines the  pairing interaction  for superconducting state with triplet pairing in centrosymmetric crystal
and has the structure corresponding to orthorhombic anisotropy in the spin space
\begin{equation}
D_{ij}({\bf k},{\bf k}') = \left(\begin{array}{ccc} D_{x}({\bf k},{\bf k}') & 0 & 0\\
0 & D_{y}({\bf k},{\bf k}') & 0 \\
0 & 0 & D_{z}({\bf k},{\bf k}')
\end{array} \right).
\end{equation}
The corresponding p-wave interaction matrix  for isotropic Fermi liquid like liquid $^3He$ is
\begin{equation}
D_{ij}^{He}({\bf k},{\bf k}')=-V({\bf k}{\bf k}')\delta_{ij}
\label{He}
\end{equation}
The explicit form of matrices of interaction in the orthorhombic ferromagnet are derived in the Appendix.

After some algebra \cite{Sam} the hamiltonian Eqn.(1) can be rewritten as hamiltonian of
pairing interaction 
\begin{equation}
H_{int}=\frac{1}{2}\sum_{{\bf k}{\bf k}'{\bf q}}V_{\alpha\beta\gamma\delta}({\bf k},{\bf k}')
a^\dagger_{{\bf k}\alpha} a^\dagger_{-{\bf k}\beta} a_{-{\bf k}'\gamma}a_{{\bf k}'\delta},
\end{equation}
here
\begin{equation}
V_{\alpha\beta\gamma\delta}({\bf k},{\bf k}')=V_{ij}({\bf k},{\bf k}')
(i\sigma_i\sigma_y)_{\alpha\delta}(i\sigma_j\sigma_y)^\dagger_{\beta\gamma}
\end{equation}
\begin{equation}
V_{ij}({\bf k},{\bf k}')=\frac{1}{2}Tr\hat D({\bf k},{\bf k}')\delta_{ij}-D_{ij}({\bf k},{\bf k}').
\end{equation}

The critical temperature is determined from the linearized self-consistency equation. In space homogenious case it has the following form \cite{Book}
\begin{eqnarray}
\Delta_{\alpha\beta}({\bf k})
=-T
\sum_{n}
\sum_{{\bf k}' }
V_{\beta\alpha\lambda\mu}({\bf k},{\bf k}')\nonumber\\
\times G_{\lambda\gamma}({\bf k}',\omega_n)
G_{\mu\delta}(-{\bf k}',-\omega_n)\Delta_{\gamma\delta}({\bf k}'),
\end{eqnarray}
where $G_{\lambda\gamma}({\bf k}',\omega_n)$
is the normal metal Green function.
The interaction functions  $D_{ij}({\bf k},{\bf k}')$ as well as the Green functions have the different form depending of the external field direction. We start from the simplest case of magnetic field parallel to the magnetization direction.

\section{Magnetic field parallel to magnetization axis}

For this field orientation the normal state  matrix Green function is
\begin{equation}
\hat G_n=\left( \begin{array}{cc}G^{{\uparrow}}& 0\\ 
0& G^{\downarrow}
\end{array}\right ),~~~~
G^{{\uparrow,\downarrow}}=\frac{1}{i\omega_n-\xi_{{\bf k}}\pm\mu_B(h+H_z)},
\end{equation}
where $h$ is the exchange field and $H_z$ is the external field.
Taking the matrix of order parameter as
\begin{equation}
\hat \Delta=\left( \begin{array}{cc}\Delta^{{\uparrow}}& \Delta^{\uparrow\downarrow}\\ 
\Delta^{\uparrow\downarrow}& \Delta^{\downarrow}
\end{array}\right )
\end{equation}
and substituting the matrices of the  Green function and the order parameter into self-consistency equation after  simple algebra we come to the system of equations
\begin{widetext}
\begin{equation}
\Delta^{\uparrow}({\bf k})
=-T
\sum_{n}
\sum_{{\bf k}' }
\left \{D^\parallel_z({\bf k},{\bf k}')
G_{1}^\uparrow
G_{2}^\uparrow
\Delta^{\uparrow}({\bf k}')
+D^\parallel_-({\bf k},{\bf k}')
G_{1}^\downarrow
G_{2}^\downarrow
\Delta^{\downarrow}({\bf k}')\right \},
\label{e1}
\end{equation}
\begin{equation}
\Delta^{\downarrow}({\bf k})
=-T
\sum_{n}
\sum_{{\bf k}' }\left \{
D^\parallel_-({\bf k},{\bf k}')
G_{1}^\uparrow
G_{2}^\uparrow
\Delta^{\uparrow}({\bf k}')
+D^\parallel_z({\bf k},{\bf k}')
G_{1}^\downarrow
G_{2}^\downarrow
\Delta^{\downarrow}({\bf k}')\right\},
\label{e2}
\end{equation}
\begin{equation}
\Delta^{\uparrow\downarrow}({\bf k})
=-\frac{1}{2}T
\sum_{n}
\sum_{{\bf k}' }\left\{
\left [ D^\parallel_+({\bf k},{\bf k}')-D^\parallel_z({\bf k},{\bf k}')\right ]
\left [G_{1}^\uparrow
G_{2}^\downarrow
+G_{1}^\downarrow
G_{2}^\uparrow
\right ]\Delta^{\uparrow\downarrow}({\bf k}')\right\}.
\end{equation}
\end{widetext}
Here we have introduced notations 
$D_\pm({\bf k},{\bf k}')=D_x({\bf k},{\bf k}')\pm D_y({\bf k},{\bf k}')$ and   $G_{1}^\uparrow=G^{\uparrow}({\bf k}',\omega_n)$, 
$G_{2}^\uparrow=G^{\uparrow}(-{\bf k}',-\omega_n)$ and similarly for the  $G_{1}^\downarrow$ and 
$G_{2}^\downarrow$ Green functions.

In the simplest case of superfluid $^3He$ with pairing interaction in the form (\ref{He}) when 
$ D_x=D_y=D_z$ we have three independent equations for $\Delta^{\uparrow}, \Delta^{\downarrow}$
and $\Delta^{\uparrow\downarrow}$.  Moreover, in absence of the exchange field $ h$ and the external field 
$H_z$  the Green functions for the  Fermi particles with up and down spin coincide each other $G^{\uparrow}=
G^{\downarrow}$. The critical temperature has the same value for all the order parameter components 
$\Delta^{\uparrow}$, $\Delta^{\downarrow}$, $\Delta^{\uparrow\downarrow}$, that corresponds to the phase transition to  one of superfluid phases of $^3He$.  The particular choice between them can be made by comparison of nonlinear terms in Ginzburg - Landau equation.
\cite{Book} In presence of the external field  the equality between the Green functions is violated
$G^{\uparrow}\ne G^{\downarrow}$.  It leads to the lifting  of the critical temperatures  degeneracy for different components of order parameter. 
As result, the highest temperature 
phase transition  occurs to the spin up-up pairing state known as $A_1$ phase of superfluid $^3He$.

In our case of orthorhombic ferromagnet the equations for order parameter components $\Delta^{\uparrow}, \Delta^{\downarrow}$, those are amplitudes of pairing of quasiparticles with spin up-up and spin down-down correspondingly,
are coupled each other. The component $\Delta^{\uparrow\downarrow}$ corresponding to pairing of quasiparticles from spin-up band with quasiparticles from spin-down band obeys the  separate equation. 
In the ferromagnetic state
in view of the large band splitting the  interband pairing superconducting state  arises  at definitely lower temperature than the spin up-up 
 and spin down-down   states.  In the paramagnetic state in general it is not true.
Here we shall not be interested in the interband pairing superconductivity.
So, we deal with system of two equations (\ref{e1}) and (\ref{e2}). 

All the interaction functions 
$D_{i}({\bf k},{\bf k}')$, here $i=x,y,z$ as one can see from Eqs.   
(\ref{D_x}),  
(\ref{D_y}) and (\ref{D_z1}) have the 
same structure
\begin{equation}
D_{i}({\bf k},{\bf k}')=a_ik_xk'_x+b_ik_yk'_y+c_ik_zk'_z.
\end{equation}
Hence,  the system of equations (\ref{e1}) and (\ref{e2}) have three independent solutions with different critical temperatures
\begin{equation}
\Delta^{\uparrow}({\bf k})=\eta^\uparrow k_x,~~~~~ \Delta^{\downarrow}({\bf k})=\eta^\downarrow k_x;
\label{X}
\end{equation}
\begin{equation}
\Delta^{\uparrow}({\bf k})=\xi^\uparrow k_y,~~~~~ \Delta^{\downarrow}({\bf k})=\xi^\downarrow k_y;
\end{equation}
\begin{equation}
\Delta^{\uparrow}({\bf k})=\zeta^\uparrow k_z,~~~~~ \Delta^{\downarrow}({\bf k})=\zeta^\downarrow k_z.
\end{equation}
Let us assume that the largest critical temperature corresponds to solution  given by Eqn. (\ref{X}).
Then performing all necessary integrations and summation in Eqs.(\ref{e1}) and (\ref{e2}) we come to the system of algebraic equations
\begin{eqnarray}
\eta^\uparrow=(g^\uparrow_\parallel\eta^\uparrow+\tilde g^{\downarrow}_\parallel\eta^\downarrow)\lambda(T)\nonumber\\
\eta^\downarrow=(\tilde g_\parallel^{\uparrow}\eta^\uparrow+g_\parallel^{\downarrow}\eta^\downarrow)\lambda(T).
\label{S}
\end{eqnarray}
According to Eqs. (\ref{D_x}),  
(\ref{D_y}) and (\ref{D_z1})
 the coefficients  in these equations are given by
\begin{equation}
g_\parallel^\uparrow=\frac{V_z\gamma_{xx}^z\langle k_x^2 N_{0}^{\uparrow}(\hat{\bf k}) \rangle}
{8\beta_z^2\left[3M_z^2-(M_z|_{H=0})^2\right]^2},
\label{1}
\end{equation}
\begin{equation}
\tilde g_\parallel^{\uparrow}=\frac{V_x\gamma_{xx}^x\langle k_x^2N_{0}^{\uparrow}(\hat{\bf k})\rangle}
{2\left[\alpha_x+2\beta_{xz}M_z^2\right]^2}-\frac{V_y\gamma_{xx}^y\langle k_x^2N_{0}^{\uparrow}(\hat{\bf k})\rangle}
{2\left[\alpha_y+2\beta_{yz}M_z^2\right]^2}.
\label{2}
\end{equation}
Here the angular brackets mean the averaging over the Fermi surface.  $N_{0}^{\uparrow}(\hat{\bf k})$
is the angular dependent density of electronic states at the Fermi surface of the band $\uparrow$.
The corresponding coefficients $g_\parallel^\downarrow$ and  $\tilde g_\parallel^\downarrow$ are obtained by the substitution $\uparrow$ to $\downarrow$ in  
Eqs. (\ref{1}) and (\ref{2}). The function $\lambda(T)$ is
\begin{equation}
\lambda(T)=2\pi T\sum_{n\ge 0}\frac{1}{\omega_n}\cong\ln\frac{\epsilon}{T},
\end{equation}
where $\epsilon$ is the energy cutoff for the pairing interaction. We assume here that it has the same value for both bands.

The
zero of the determinant of  the system (\ref{S}) yields BCS like formula
\begin{equation}
T=\epsilon\exp\left (-\frac{1}{g_{\parallel}}\right ),
\label{CT}
\end{equation}
where, the "constant" of interaction
\begin{equation}
g_{\parallel}=\frac{g_\parallel^\uparrow+g_\parallel^\downarrow}{2}+\sqrt{\frac{(g_\parallel^\uparrow-g_\parallel^\downarrow)^2}{4}+
\tilde g_\parallel^\uparrow\tilde g_\parallel^\downarrow}
\end{equation}
is the function of temperature and magnetic field. Thus,  the formula (\ref{CT}) is in fact the equation for determination of the critical temperature of the transition to the superconductig state
$T_{sc}=T_{sc}(H)$.

Let us first consider the situation at $H_z=0$. Taking into account expressions (\ref{1}) and (\ref{2}) the equation (\ref{CT}) can be rewritten as 
\begin{equation}
\ln\frac{T}{\epsilon}=-\frac{1}{g_\parallel}=-(T-T_c)^2f(T),
\label{CTT}
\end{equation}
where $f(T)\cong const$ is slowly varying function of temperature. There are three different situations.

(i) The Curie temperature is much larger than the cutoff energy, that is the band splitting is large in comparison with the width of region of effectiveness of pairing interaction. This case the solution of Eq. (\ref{CTT}) obeys inequality
\begin{equation}
T_{sc}\ll\epsilon\ll T_c.
\end{equation}
An increasing of the Curie temperature say as a function of pressure shifts to the right the parabola
in the right hand side of Eq. (\ref{CTT}). Hence, its intersection with logarithm in the left hand side  is shifted to lower temperature.
Thus, the  critical temperature of transition to the superconducting state decreases with the Curie temperature increasing as it is indeed the case in $URhGe$. \cite{Hardy}  The opposite tendency
takes place at decreasing of the Curie temperature accompanying by the increasing of the superconducting critical temperature as it is in $UCoGe$.\cite{Slooten,Hassinger}  

(ii) The outlined above behavior, however, has no proper description in the region where the  Curie temperature decreases approching to the cutoff energy. This case the formal solution of Eq. (\ref{CTT}) is $T_{sc}=\epsilon=T_c$.
It means that we are out   applicability of BCS theory. So, the proper treatment of the problem in vicinity of intersection of the $T_c(P)$ and $T_{sc}(P)$ curves is out of frame of applicability of developed approach. 

(iii) Finally one can consider the situation deeply in the paramagnetic region where the solution of Eq.(\ref{CTT}) obeys inequality 
\begin{equation}
T_c\ll T_{sc}\ll\epsilon.
\end{equation}
Here, a displacement of parabola to the left shifts solution of Eq. (\ref{CTT}) to the left.
Hence, the Curie temperature decrease with pressure is accompanied by the decreasing of the temperature of superconducting transition. 

Let us turn now to the situation with $H_z\ne 0$. In the ferromagnetic state the external field in the direction of the spontaneous magnetization drastically suppresses the superconducting state. Indeed,  on the one hand the field dependence of the band density of states leads to the tiny  shift in critical temperatue $\delta T_{sc}/T_{sc}\sim\pm \frac{\mu_BH_z}{\varepsilon_F}\ln \frac{\varepsilon_F}{T_{sc}}$.\cite{Min04} On the other hand,  the increase of the magnetization under the influence of the parallel external field effectively suppress the coupling "constants".  For instance,  at temperatures well below the Curie temperature the zero field magnetization is almost temperature independent.  And the right hand side of formula (\ref{CT}) is temperature independent as well. Then the critical temperature field dependence
\begin{equation}
T_{sc}(H_z)=\epsilon\exp\left (-\frac{1}{g_{\parallel}(H_z)}\right )
\label{CTH}
\end{equation}
 is determined by the field dependence of "constants" of interaction $g_\parallel^\uparrow,
\tilde g_\parallel^{\uparrow},...$ which are expressed through the field dependent magnitude of magnetization.
The magnetic moment $M_z^0$ under magnetic field $\sim0.5$ Tesla, which is of the order of the upper critical field at zero temperature,   acquires $\sim1.3$ times increase  in respect to its zero field value\cite{Huy08}. According to Eq. (\ref{1}) it
leads to the  
4 times decreasing of the coupling constants  $g^\uparrow$ and $g^\downarrow$  !  The decreasing of the constants $\tilde g^\uparrow$ and $\tilde g^\downarrow$  (see Eq. (\ref{2})) is not so impressive, but also certainly takes place.    So, the magnetic field parallel to the spontaneous magnetization effectively suppresses the pairing in the ferromagnetic state.

 In the paramagnetic state   
 in presence of the external field the coefficients $ g_\uparrow\ne g_\downarrow$
 and $ \tilde g_\uparrow\ne\tilde g_\downarrow$ and given by formulae
 \begin{equation}
g^\uparrow=\frac{V_z\gamma_{xx}^z\langle k_x^2 N_{0\uparrow}(\hat{\bf k}) \rangle}
{2\left[\alpha_z+6\beta_zM_z^2\right]^2},
\label{3}
\end{equation}
\begin{equation}
\tilde g^{\uparrow}=\frac{V_x\gamma_{xx}^x\langle k_x^2N_{0\uparrow}(\hat{\bf k})\rangle}
{2\left[\alpha_x+2\beta_{xz}M_z^2\right]^2}-\frac{V_y\gamma_{xx}^y\langle k_x^2N_{0\uparrow}(\hat{\bf k})\rangle}
{2\left[\alpha_y+2\beta_{yz}M_z^2\right]^2}.
\label{4}
\end{equation}
 where the magnetization has the field proportional value (see Eq.(\ref{M_zp})). Thus, above the Curie temperature  the external magnetic field along the easiest magnetization axis suppresses  the superconducting state as well  as it does in ferromagnetic state for the field orientation parallel to spontaneous magnetization.

\section{Magnetic field perpendicular to magnetization axis}

For the field  orientated in a perpendicular to the spontaneous magnetization direction say along  b-axis 
 it is natural to choose the spin quantization axis along the  direction of the total magnetic field $
 h\hat z+H_y\hat y$. Then the normal state  matrix Green function is  diagonal 
\begin{equation}
\hat G_n=\left( \begin{array}{cc}G^{{\uparrow}}& 0\\ 
0 & G^{\downarrow}
\end{array}\right ),
\end{equation}
where
\begin{equation}
G_{{\uparrow,\downarrow}}=\frac{1}{i\omega_n-\xi_{{\bf k}}\pm\mu_B\sqrt{h^2+H_y^2}
}.
\end{equation}
The potential of interaction found  in the Appendix should be rewritten in the new coordinate frame
\begin{equation}
V_{ij}({\bf k},{\bf k}')=R_{il}\left (\frac{1}{2}Tr\hat D^\perp({\bf k},{\bf k}')\delta_{lm}-D^\perp_{lm}({\bf k},{\bf k}'\right)\tilde R_{mj},
\end{equation}
where 
\begin{equation}
\hat R=\left( \begin{array}{ccc}1& 0 & 0\\ 
0 &\cos\varphi&-\sin\varphi\\
0&\sin\varphi&  \cos\varphi
\end{array}\right ),
\end{equation}
is matrix of rotation around $\hat x$ direction on the angle given by 
$
\tan\varphi={H_y}/{h}
$, and $\tilde R_{mj}$ is the transposed matrix.

The system of self-consistency equations acquires the following form
\begin{widetext}
\begin{eqnarray}
&\Delta^{\uparrow}({\bf k})
=-T
\sum_{n}
\sum_{{\bf k}' }
\left\{\left[D^\perp_z({\bf k},{\bf k}')\cos^2\varphi+D^\perp_y({\bf k},{\bf k}')\sin^2\varphi\right]
G_{1}^\uparrow
G_{2}^\uparrow
\Delta^{\uparrow}({\bf k}')\right.\nonumber\\
&+\left.\left[D^\perp_-({\bf k},{\bf k}')\cos^2\varphi+(D^\perp_x({\bf k},{\bf k}')-D^\perp_z({\bf k},{\bf k}'))\sin^2\varphi\right]
G_{1}^\downarrow
G_{2}^\downarrow
\Delta^{\downarrow}({\bf k}')\right.\nonumber\\
&\left.+i\left [D^\perp_z({\bf k},{\bf k}')-D^\perp_y({\bf k},{\bf k}')\right]\sin\varphi\cos\varphi\left(G_1^\uparrow G_2^\downarrow +G_1^\downarrow G_2^\uparrow\right)
\Delta^{\uparrow\downarrow}({\bf k}')\right\},
\label{e1'}
\end{eqnarray}
\begin{eqnarray}
&\Delta^{\downarrow}({\bf k})
=-T
\sum_{n}
\sum_{{\bf k}' }
\left\{\left[D^\perp_-({\bf k},{\bf k}')\cos^2\varphi+(D^\perp_x({\bf k},{\bf k}')-D^\perp_z({\bf k},{\bf k}'))\sin^2\varphi\right]
G_{1}^\uparrow
G_{2}^\uparrow
\Delta^{\uparrow}({\bf k}')\right.\nonumber\\
&+\left.\left[D^\perp_z({\bf k},{\bf k}')\cos^2\varphi+D^\perp_y({\bf k},{\bf k}')\sin^2\varphi\right]
G_{1}^\downarrow
G_{2}^\downarrow
\Delta^{\downarrow}({\bf k}')\right.\nonumber\\
&\left.+i\left [D^\perp_z({\bf k},{\bf k}')-D^\perp_y({\bf k},{\bf k}')\right]\sin\varphi\cos\varphi\left(G_1^\uparrow G_2^\downarrow +G_1^\downarrow G_2^\uparrow\right)
\Delta^{\uparrow\downarrow}({\bf k}')\right\},
\label{e2'}
\end{eqnarray}
\begin{eqnarray}
&\Delta^{\uparrow\downarrow}({\bf k})
=-\frac{1}{2}T
\sum_{n}
\sum_{{\bf k}' }\left\{
-i\left[D^\perp_z({\bf k},{\bf k}')-D^\perp_y({\bf k},{\bf k}')\right]\sin2\varphi
\left[G_{1}^\uparrow
G_{2}^\uparrow
\Delta^{\uparrow}({\bf k}')+G_{1}^\downarrow
G_{2}^\downarrow
\Delta^{\downarrow}({\bf k}') \right]
\right.
\nonumber\\
&\left.+\left[(D^\perp_+({\bf k},{\bf k}')-D^\perp_z({\bf k},{\bf k}'))\cos^2\varphi+(D^\perp_-({\bf k},{\bf k}')+D^\perp_z({\bf k},{\bf k}'))
\sin^2\varphi \right]\left(G_1^\uparrow G_2^\downarrow +G_1^\downarrow G_2^\uparrow
\right)
\Delta^{\uparrow\downarrow}({\bf k}')\right\}.
\label{e3'}
\end{eqnarray}
\end{widetext}
Unlike to the case of parallel  field here all the components of the order parameter are coupled each other. In principle one can find  the solution of the whole system.  However, in view of large band splitting one may neglect the last term containing the products $G_1^\uparrow G_2^\downarrow +G_1^\downarrow G_2^\uparrow$ in the equations (\ref{e1'}) and  (\ref{e2'}).
Then the equations for the$ \Delta^\uparrow$ and $\Delta^\downarrow$ are decoupled from the equation (\ref{e3'}).  The system acquires the same form 
\begin{widetext}
\begin{eqnarray}
&\Delta^{\uparrow}({\bf k})
=-T
\sum_{n}
\sum_{{\bf k}' }
\left\{\left[D^\perp_z({\bf k},{\bf k}')\cos^2\varphi+D^\perp_y({\bf k},{\bf k}')\sin^2\varphi\right]
G_{1}^\uparrow
G_{2}^\uparrow
\Delta^{\uparrow}({\bf k}')\right.\nonumber\\
&+\left.\left[D^\perp_-({\bf k},{\bf k}')\cos^2\varphi+(D^\perp_x({\bf k},{\bf k}')-D^\perp_z({\bf k},{\bf k}'))\sin^2\varphi\right]
G_{1}^\downarrow
G_{2}^\downarrow
\Delta^{\downarrow}({\bf k}')
\right\},
\label{e11}
\end{eqnarray}
\begin{eqnarray}
&\Delta^{\downarrow}({\bf k})
=-T
\sum_{n}
\sum_{{\bf k}' }
\left\{\left[D^\perp_-({\bf k},{\bf k}')\cos^2\varphi+(D^\perp_x({\bf k},{\bf k}')-D^\perp_z({\bf k},{\bf k}'))\sin^2\varphi\right]
G_{1}^\uparrow
G_{2}^\uparrow
\Delta^{\uparrow}({\bf k}')\right.\nonumber\\
&+\left.\left[D^\perp_z({\bf k},{\bf k}')\cos^2\varphi+D^\perp_y({\bf k},{\bf k}')\sin^2\varphi\right]
G_{1}^\downarrow
G_{2}^\downarrow
\Delta^{\downarrow}({\bf k}')
\right\},
\label{e21}
\end{eqnarray}
\end{widetext}
as it had in  the case of parallel field given by Eqs. (\ref{e1}) and (\ref{e2}).
Hence, assuming again that the highest critical temperature corresponds to solution  given by Eqn. 
(\ref{X}) and performing all necessary integrations and summation in Eqs. (\ref{e11}) and (\ref{e21}) we come to the system of algebraic equations
\begin{eqnarray}
\eta^\uparrow=(g_\perp^\uparrow\eta^\uparrow+\tilde g^{\downarrow}_\perp\eta^\downarrow)\lambda(T)\nonumber\\
\eta^\downarrow=(\tilde g_\perp^{\uparrow}\eta^\uparrow+g_\perp^{\downarrow}\eta^\downarrow)\lambda(T).
\label{S''}
\end{eqnarray}
According to Eqs. (\ref{D_z11}),  
(\ref{D_x11}),  
(\ref{D_y11})  the coefficients  in these equations are given by
\begin{equation}
g_\perp^\uparrow=\frac{V_z\gamma_{xx}^z\langle k_x^2 N_{0}^{\uparrow}(\hat{\bf k}) \rangle}
{32\beta_z^2M_z^4}\cos^2\varphi+\frac{V_y\gamma_{xx}^y\langle k_x^2N_{0}^{\uparrow}(\hat{\bf k})\rangle}
{2\left[\alpha_y+2\beta_{yz}M_z^2\right]^2}\sin^2\varphi,
\label{111}
\end{equation}
\begin{eqnarray}
\tilde g_\perp^{\uparrow}=\left(\frac{V_x\gamma_{xx}^x\langle k_x^2N_{0}^{\uparrow}(\hat{\bf k})\rangle}
{2\left[\alpha_x+2\beta_{xz}M_z^2\right]^2}-\frac{V_y\gamma_{xx}^y\langle k_x^2N_{0}^{\uparrow}(\hat{\bf k})\rangle}
{2\left[\alpha_y+2\beta_{yz}M_z^2\right]^2}\right)\cos^2\varphi\nonumber\\
+\left (\frac{V_x\gamma_{xx}^x\langle k_x^2N_{0}^{\uparrow}(\hat{\bf k})\rangle}
{2\left[\alpha_x+2\beta_{xz}M_z^2\right]^2}- \frac{V_z\gamma_{xx}^z\langle k_x^2 N_{0}^{\uparrow}(\hat{\bf k}) \rangle}
{32\beta_z^2M_z^4}  \right)\sin^2\varphi~~
\label{222}
\end{eqnarray}
The corresponding coefficients $g_\parallel^\downarrow$ and  $\tilde g_\parallel^\downarrow$ are obtained by the substitution $\uparrow$ to $\downarrow$ in  
Eqs. (\ref{111}) and (\ref{222}). 

The
zero of the determinant of  the system (\ref{S}) yields BCS like formula
\begin{equation}
T=\epsilon\exp\left (-\frac{1}{g_{\perp}}\right ),
\label{CT1}
\end{equation}
where, the "constant" of interaction
\begin{equation}
g_{\perp}=\frac{g_\perp^\uparrow+g_\perp^\downarrow}{2}+\sqrt{\frac{(g_\perp^\uparrow-g_\perp^\downarrow)^2}{4}+
\tilde g_\perp^\uparrow\tilde g_\perp^\downarrow}
\end{equation}
is the function of temperature and magnetic field. 

The easiest way to follow up the critical temperature field dependence  is to consider temperature region
  well below the Curie temperature where the zero field magnetization is almost temperature independent.  And the right hand side of formula (\ref{CT1}) is temperature independent as well. Then the critical temperature field dependence
\begin{equation}
T_{sc}(H_y)=\epsilon\exp\left (-\frac{1}{g_{\perp}(H_y)}\right )
\label{CTH1}
\end{equation}
 is determined by the field dependence of "constants" of interaction $g_\perp^\uparrow,
\tilde g_\perp^{\uparrow},...$  originating  from the field dependent  trigonometric factors 
\begin{equation}
\cos^2\varphi=\frac{h^2}{h^2+H_y^2},~~~~~~~ \sin^2\varphi=\frac{H_y^2}{h^2+H_y^2}
\end{equation}
and the field dependent magnitude of magnetization $M_z(H_y)$ to be determined from Eqs (\ref{A21}) and (\ref{A22}). 
 
The trigonometric factors are changed with $H_y$ on the scale of exchange field. Hence, if  the $M_z(H_y)$ is just slightly decreased at $H_y\sim H_{c2y}\ll h$, as it is in URhGe in the region of moderate fields \cite{Hardy05,Levy05}, where $M_z(H_y=0)$ is big enough, then  the critical temperature slowly decreases due to field dependence of trigonometric factors;
 \begin{equation}
 \frac{T_{sc}(H_y)-T_{sc}(0)}{T_{sc}(0)}\propto -\frac{H_y^2}{h^2}
 \end{equation}
Obviously, this corresponds to 
small paramagnetic suppression of superconductivity have been considered in the paper \cite{Min10}.

Another situation takes place in UCoGe. Here the magnitude of zero field magnetization $M_z(H_y=0)$ is significantly smaller. Apparently, the field induced decrease of $M_z(H_y)$ in this material   is faster \cite{footnote}, 
that leads to the effective  increasing the "constants" of interaction $g^\uparrow$, $g^\downarrow$ despite of its decrease due to the trigonometric factors.
This mechanism explains the effect of upward curvature in the temperature dependence of the upper critical fields in a  and b directions observed in UCoGe \cite{Huy08,Slooten,Aoki09}. At the same time  the phenomenon of  S shape temperature dependence of the upper critical field in b-direction apparently related with significant increase of effective mass \cite{Aoki09} observed in field interval  from 5 to 15 Tesla. Out this  interval 
for field  in $b$-direction and for any field in $a$-direction the average value of effective mass decreases with field increase and cannot be responsible for the effect of the stimulation of superconductivity.

\section{Conclusion}

We have studied  the magnetic field dependence of pairing interaction induced by magnetic fluctuations in the ferromagnetic superconducting compounds URhGe and UCoGe with orthorhombic crystal structure. 

For the field orientation along the spontaneous magnetization there was demonstrated the effect of suppression of fluctuations and, hence, of  the superconductivity by magnetic field.   The leading role plays here the  suppression of longitudinal fluctuations determining   the constants $g^\uparrow$ and $g^\downarrow$ given by Eq.(\ref{1}), that corresponds  strongly anisotropic Izing like magnetism in  uranium compounds \cite{Huy08,Ihara10}.

For the field directed perpendicular to the spontaneous magnetization the field dependence of the constants of interaction is determined by interplay of paramagnetic suppression of superconductivity which is significantly weakened in the ferromagnetic superconductors \cite{Min10} and the effect of stimulation of superconductivity by means the increasing of longitudinal  magnetic fluctuations  magnetic field given by Eq. (\ref{111}).

\acknowledgments

This work was partly supported by the grant SINUS of  the Agence Nationale de la Recherche.

The author is indebted to J.-P. Brison and A. de Visser for  stimulating questions concerning the peculiar behavior of the upper critical field in ferromagnetic superconductors. I am also thankful  to K. Ishida, D. Aoki, F. Hardy and J. Flouquet  who have made me familiar with their recent experimental results, and to M. Zhitomirsky for the
enlightening discussions.

\appendix

\section{Magnetic susceptibility of orthorhombic ferromagnet}

URhGe and UCoGe  are the orthorhombic ferromagnets with spontaneous magnetization oriented along $c$ crystallography axis. At the temperatures below the Curie
temperature and in the absence of magnetic field
the $c$ component of magnetization has a finite value.  Here we derive  the magnetic susceptibility
of a ferromagnet with orthorhombic symmetry. It  determines  the pairing interactions due to spin fluctuations in such  type of materials. To begin  we write
the Landau free energy of orthorhombic ferromagnet in
magnetic field ${\bf H}({\bf r})$
\begin{equation}
{\cal F}=\int d V(F_M+F_\nabla),
\label{FE}
\end{equation}
where 
\begin{eqnarray}
F_M=\alpha_{z}M_{z}¥^{2}+\alpha_{y}M_{y}^{2}+\alpha_{x}M_{x}¥^{2}~~~~~~~~~~~~~~~~~\nonumber\\
+\beta_{z}¥M_{z}¥^{4} +\beta_{yz}¥M_{z}¥^{2}¥M_{y}¥^{2}¥+\beta_{xz}¥M_{z}¥^{2}¥M_{x}¥^{2}-{\bf M}{\bf  H},
\label{F}
\end{eqnarray}
 and
\begin{equation}
F_\nabla=\gamma_{ij}^x\frac{\partial M_x}{\partial x_i}\frac{\partial M_x}{\partial x_j}
+\gamma_{ij}^y\frac{\partial M_y}{\partial x_i}\frac{\partial M_y}{\partial x_j}
+\gamma_{ij}^z\frac{\partial M_z}{\partial x_i}\frac{\partial M_z}{\partial x_j}
\label{F_{nabla}}
\end{equation}
is the density of gradient energy.
Here the $x, y, z$ are directions of the spin axes pinned to $a, b, c$
crystallographic directions correspondingly, 
\begin{equation}
\alpha_{z}=\alpha_{z0}¥(T-T_c), 
\end{equation}
$\alpha_x>0$, $\alpha_y>0$ and matrices $\gamma_{ij}^p$, where $p=x,y,z$, have the form
\begin{equation}
\gamma_{ij}^p = \left(\begin{array}{ccc} \gamma_{xx}^p & 0 & 0\\
0 & \gamma_{yy}^p & 0 \\
0 & 0 & \gamma_{yy}^p
\end{array} \right).
\end{equation}
Starting this point we shall separately consider the  ferromagnet susceptibility under stationary magnetic field  parallel either to the magnetization direction or parrallel to  one of the perpendicular crystallographic directions. \cite{El}

\subsection{ Magnetic field parallel to magnetization axis}

So, let us take the magnetic field in the form
\begin{equation}
{\bf H}({\bf r})=H_z\hat z+\delta{\bf H}({\bf r}),
\end{equation}
where $|\delta{\bf H}({\bf r})|\ll H_z$.
By variation of the functional Eq.(\ref{FE}) in respect to $M_z$ we come to 
\begin{equation}
2\alpha_zM_z+4\beta_zM_z^3+2\beta_{yz}M_zM_y^2-\gamma_{ij}^z\frac{\partial^2 M_z}{\partial x_i
\partial x_j}-H_z-\delta H_z({\bf r})=0
\end{equation}
We shall search for the solution of this equation
as
\begin{equation}
M_z({\bf r})=M_z+\delta M_z({\bf r}),
\end{equation}
where $M_z$  is the solution of the space homogeneous  problem
\begin{equation}
2\alpha_zM_z+4\beta_zM_z^3-H_z=0.
\label{eq}
\end{equation}
Here we omit the term $2\beta_{yz}M_zM_y^2$ in view of zero 
coordinate independent value of  magnetization  
$$M_y=0$$ in $y$ direction.
In absence of field the magnetization takes value
\begin{equation}
(M_z|_{H=0})^2=-\frac{\alpha_z}{2\beta}.
\label{M_z}
\end{equation}
For the $\delta M_z$ we have equation
\begin{equation}
2\alpha_z\delta M_z+12\beta_zM_z^2\delta M_z-\gamma_{ij}^z\frac{\partial^2 \delta M_z}{\partial x_i
\partial x_j}=\delta H_z({\bf r})
\end{equation}
The solution of this equation in the ${\bf k}$ space is
\begin{equation}
\delta M_{z}({\bf k})=\chi_z({\bf k})\delta H_z({\bf k}),
\end{equation}
where
\begin{equation}
\chi_z({\bf k})=\frac{1}{2\alpha_z+12\beta_zM_z^2+\gamma_{ij}^zk_ik_j}
\end{equation}
The function determining the pairing interaction  in triplet channel \cite{Sam} is obtained from the subseptibility
$\chi_z({\bf k}-{\bf k}')$ as its part which is odd function of both of its arguments
\begin{eqnarray}
D^\parallel_z({\bf k},{\bf k}')=-
\frac{V_z}{2}[\chi_z({\bf k}-{\bf k}')-\chi_z({\bf k}+{\bf k}')]\nonumber\\
\approx -\frac{V_z\gamma_{ij}^zk_ik'_j}
{2\left[\alpha_z+6\beta_zM_z^2\right]^2},
\label{D_z}
\end{eqnarray}
where $V_z>0$ is the constant of interaction.

In similar manner one can obtain the expressions for the interaction functions in two  other directions
\begin{equation}
D^\parallel_x({\bf k},{\bf k}')
\approx-\frac{V_x\gamma_{ij}^xk_ik'_j}
{2\left[\alpha_x+2\beta_{xz}M_z^2\right]^2}
\label{D_x}
\end{equation}
and 
\begin{equation}
D^\parallel_y({\bf k},{\bf k}')
\approx-\frac{V_y\gamma_{ij}^yk_ik'_j}
{2\left[\alpha_y+2\beta_{yz}M_z^2\right]^2},
\label{D_y}
\end{equation}
where we have  introduced the constants of interaction $V_x, V_y$. 
Taking into account Eq. (\ref{M_z}) one can rewrite
the
expression for $D_z({\bf k},{\bf k}')$ as follows
\begin{equation}
D^\parallel_z({\bf k},{\bf k}')= -\frac{V_z\gamma_{ij}^zk_ik'_j}
{8\beta_z^2\left[3M_z^2-(M_z|_{H=0})^2\right]^2}.
\label{D_z1}
\end{equation}

In the paramagnetic state the functions $D_z({\bf k},{\bf k}')$, $D_x({\bf k},{\bf k}')$ and $D_y({\bf k},{\bf k}')$ are given by the same Eqs. (\ref{D_z}), (\ref{D_x}), (\ref{D_y}) where
the equilibrium value of magnetization is proportional to the external field
\begin{equation}
M_z=\frac{H_z}{2\alpha_z}.
\label{M_zp}
\end{equation}
Here, one must take in mind that the magnetization growing up at Curie temperature when $\alpha_z(T)\to 0$ is limited by the nonlinear term in Eq. (\ref{eq}).

\subsection{ Magnetic field perpendicular to magnetization axis}

So, let us take the magnetic field in the form
\begin{equation}
{\bf H}({\bf r})=H_y\hat y+\delta{\bf H}({\bf r}),
\end{equation}
where $|\delta{\bf H}({\bf r})|\ll H_y$.
Stationary vaue of magnetization is determined by the system of equations
\begin{equation}
2\alpha_yM_y+2\beta_{yz}M_z^2M_y-H_y=0,
\end{equation}
\begin{equation}
\alpha_z+2\beta_zM_z^2+\beta_{yz}M_y^2=0.
\label{A21}
\end{equation}

The field induced stationary magnetization along b-direction is
\begin{equation}
M_y=\frac{H_y}{2(\alpha_y+\beta_{yz}M_z^2)}.
\label{A22}
\end{equation} 
Substituting this value back in the eqn. (\ref{F}) we obtain at $\beta_{yz}M_z^2/\alpha_y<1$, that is certainly true  not so far from the Curie temperature,
\begin{equation}
F=\alpha_{0z}¥\left(T-T_c+\frac{\beta_{yz}H_y^2}{4\alpha_{z0}\alpha_y^2}\right)M_{z}¥^{2}¥+ \beta_{z}M_{z}¥^{4}.
\label{fe}
\end{equation}
Hence, the Curie temperature 
\begin{equation}
T_{Curie}(H_y)=T_c-\frac{\beta_{yz}H_y^2}{4\alpha_{z0}\alpha_y^2}
\end{equation} 
is suppressed by the magnetic field oriented along b-axis.
This type of behavior was observed in UCoGe.\cite{Aoki09} 
The magnetization along z-direction is also decreased
\begin{equation}
M_z^2=\frac{\alpha_{z0}(T_c-T)}{2\beta_z}-\frac{\beta_{yz}H_y^2}{8\alpha_y^2\beta_z}
\end{equation}
Far from the Curie temperature, where this particular formula is not valid, the decrease of 
$z$-component of magnetization determined by Eqs. (\ref{A21}) and (\ref{A22})
with growth  $H_y$ still takes place. 

Following procedure of the 
previous section of Appendix we obtain the susceptibility along $z$ didection
\begin{equation}
\chi_z({\bf k})=\frac{1}{2\alpha_z+12\beta_zM_z^2+2\beta_{yz}M^2_y+\gamma_{ij}^zk_ik_j},
\end{equation}
that making use Eq.(\ref{A21}) can be rewritten as
\begin{equation}
\chi_z({\bf k})=\frac{1}{8\beta_zM_z^2+\gamma_{ij}^zk_ik_j},
\end{equation}
The corresponding function determining the pairing interaction is
\begin{equation}
D^\perp_z({\bf k},{\bf k}')
\approx -\frac{V_z\gamma_{ij}^zk_ik'_j}
{32\beta_z^2M_z^4},
\label{D_z11}
\end{equation}
By comparison with Eq.(\ref{D_z1}) in the absence of field  we  naturally come to $D^\perp_z=D^\parallel_z$.

The expressions for the interaction functions in two  other directions are 
\begin{equation}
D^\perp_x({\bf k},{\bf k}')
\approx-\frac{V_x\gamma_{ij}^xk_ik'_j}
{2\left[\alpha_x+2\beta_{xz}M_z^2\right]^2}
\label{D_x11}
\end{equation}
and 
\begin{equation}
D^\perp_y({\bf k},{\bf k}')
\approx-\frac{V_y\gamma_{ij}^yk_ik'_j}
{2\left[\alpha_y+2\beta_{yz}M_z^2\right]^2}.
\label{D_y11}
\end{equation}
They have the same form as for the external field directed along the spontaneous magnetization. 
However, for the field orientation perpendicular to the spontaneous magnetization $M_z$ decreases with field increase that leads to the growth the pairing interaction and increase the temperature of superconducting phase transition.


\begin{thebibliography}{99}
\bibitem{footnote} The corresponding experimental data at the moment are absent.
\end{thebibliography}
\end{document}